\documentclass[12pt]{article}
\usepackage[dvips]{graphicx}
\def\frac#1#2{{#1\over #2}}
\def\lsim{\mathrel{\rlap{\lower4pt\hbox{\hskip1pt$\sim$}}
    \raise1pt\hbox{$<$}}}         
\def\gsim{\mathrel{\rlap{\lower4pt\hbox{\hskip1pt$\sim$}}
    \raise1pt\hbox{$>$}}}         

\def\Imag#1{\Im{\rm m}#1}
\def\Real#1{\Re{\rm e}#1}

\def\ie{\hbox{\it i.e. }}
\def\etc{\hbox{\it etc... }}
\def\eg{\hbox{\it e.g. }}

\def\t{$|t|$ }

\begin{document}

\rightline{LYCEN 99132}

\rightline{ DFTT 65/99}

\bigskip
\begin{center}

{\Large \bf Of DIPS, STRUCTURES and EIKONALIZATION}
\bigskip
\bigskip

{\bf P. Desgrolard}({\footnote{E-mail: desgrolard@ipnl.in2p3.fr}}),
{\bf M. Giffon}({\footnote{E-mail: giffon@ipnl.in2p3.fr}}),
{\bf E. Martynov}({\footnote{E-mail: martynov@bitp.kiev.ua}}),
{\bf E. Predazzi}({\footnote{ E-mail: predazzi@to.infn.it}}).

\end{center}

\bigskip
$^{1,2}$){\it Institut de Physique Nucl\'eaire de Lyon, IN2P3-CNRS et
Universit\'{e}\\ Claude Bernard,\\ 43 boulevard du 11 novembre 1918,
F-69622 Villeurbanne Cedex, France\\}

($^3$){\it Bogoliubov Institute for Theoretical Physics, National
Academy of Sciences of Ukraine, 03143, Kiev-143, Metrologicheskaja
14b, Ukraine\\}

($^4$){\it
Dipartimento di Fisica Teorica - Universit\`a di Torino
and Sezione INFN di Torino, Italy}

\bigskip
{\bf Abstract}

We have investigated several models of Pomeron and Odderon
contributions to high energy elastic $pp$ and $\bar p p$ scattering.
The questions we address concern their role in this field, the
behavior of the scattering amplitude (or of the
total cross-section) at
high energy, and how to fit all high energy elastic data. The data
are extremely well reproduced by our approach at all momenta and for
sufficiently high energies. The relative virtues of Born amplitudes
and of different kinds of eikonalizations are considered. An
important point in this respect is that secondary structures are
predicted in the differential cross-sections at increasing energies
and these phenomena appear quite directly related to the procedure of
eikonalizing the various Born amplitudes. We conclude that these
secondary structures arise naturally within the eikonalized procedure
(although their precise localization turns out to be model
dependent). The fitting procedure naturally predicts the appearance
of a zero at small $|t|$ in the real part of the even amplitude as
anticipated by general theorems. We would like to stress, once again,
how important it would be to have at LHC both $pp$ and $p \bar p$
options for
many questions connected to the general properties of high
energy hadronic physics
and for a check of our predictions.

\bigskip

\section{\bf Introduction}
Few years ago~\cite{dgp},
combining several of the currently used philosophies, a high quality
description of existing high energy elastic $pp$ and $\bar pp$
scattering data was obtained. The main lessons of this study
performed at the Born level were:

{\it (1)} an Odderon contribution is absolutely necessary to
reproduce quantitatively well the data; while its presence is not
explicitly needed at $t=0$, its inclusion is necessary to have a good
fit of the other \t data, specially in the dip region and in the
high-\t domain

{\it (2)} hints are found that secondary structures
(diffraction-like) develop in angular distributions with increasing
energies at intermediate \t values in both $pp$ and
$p\bar p$ angular distributions.

In particular, it was suggested that
such structure effects should be well visible at LHC while only extremely
precise data could perhaps show the effect at RHIC energies.

However, answers to some important points are still incomplete.
In particular, what is a good model for the Pomeron~? What is the
behavior of the scattering amplitude at high ("asymptotic") energy~?
Are large-\t data dominated by the Odderon~? Better, does a special
criterium exist proving the Odderon presence~? Can one settle the
question about the sign of $\alpha_O(0)-1\equiv\delta_O$ concerning
the intercept{\footnote {Originally~\cite{gln} it was claimed that
$\delta_O>0$. More recently, counterarguments have been
given~\cite{wos} to suggest that $\delta_O$ should be negative. This
possibility had been anticipated in~\cite{dgp}
on purely phenomenological grounds and, subsequently, we have found that
such a requirement is, in general, consequence of unitarity ~\cite{eikge}.

However latest QCD calculation~\cite{bartels} gives $\delta_O=0$.} }
of the Odderon~?
Are secondary structures always predicted at large-\t when $s$
increases, \ie do they arise "naturally" and are they model
dependent~? What is the r\^ole of eikonalization~? Does an
amplitude that fits well the data exhibit automatically a zero in the
real part of the even component of the amplitude as required by
general theorems~\cite{martin}?

Only partial answers presently exist (see~\cite{blois95} and references
therein).

The Pomeron remains a most mysterious entity in spite of its
resurgence from Diffractive Deep Inelastic Scattering (DDIS)
data{\footnote{For an update on the subject, see \eg~\cite{el,ep}.}}.
Many models, however, exist and we are going to probe a few. Even at
low-\t (first diffraction cone), it has been shown~\cite{dlm} that
existing data do not allow to select among Pomeron models. The
present data are, very likely, not yet asymptotic; this
(see~\cite{dglm} and references therein) makes it very difficult with
the existing data to establish a definite asymptotic behavior for the
amplitude.

The r\^ole of eikonalization has not been fully clarified in spite of
having been investigated by many authors~\cite{ter} but many results
have been obtained recently~\cite{gmp,eikge}.

The Odderon is instrumental in reproducing the large-\t data. While
$t=0$ da\-ta are presumably dominated by the Pomeron, which in this
region hides the Odderon, very precise data could be useful to shed
light on its existence~\cite{gps}.

Predictions of secondary structures have appeared many times in the
past~\cite{past}. The large spectrum of predictions in the position
of these secondary dips shows that things are actually more
complicated than anticipated long ago~\cite{hz}. It is not enough
that a given scheme inherently generates oscillations (like the
Bessel function of an impact parameter representation); interference
effects are very important in determining their position. The model
dependence of these predictions, however, is not so important; it is
the prediction itself of the existence of secondary structures which
matters.

In this paper, four of the above points are taken into special
considerations. The first is the investigation of the r\^ole and
properties of the different varieties of
eikonalization procedures one can devise. The second concerns the
appearance of secondary dips and structures. These two points are
strictly interconnected, being the second, to some extent, the
physical counterpart of the other. The third is devoted to the
behavior of the real part of the even amplitude close to zero. The
fourth concerns the r\^ole of the Odderon in the construction of the
amplitude and in the reproduction of the data.

The eikonalization procedure and its consequences is one of the
principal subjects we discuss in this paper. We briefly revise (in
Sec.~3) the {\it Ordinary Eikonalization} (OE) and, after
(re)discovering its limits, we proceed to discuss a one-parameter
generalization called {\it Quasi Eikonalization} (QE)~\cite{ter} and
to propose a three-parameter extension which we term {\it Generalized
Eikonalization} (GE) (see~\cite{eikge}). Although a useful tool to
alleviate the violations of $s$-channel unitarity at some level (as
emphasized in~\cite{gmp}), eikonalization does not mean
unitarization.

The effects of Ordinary Eikonalization
as compared to the use of the Born amplitude
have been studied within
a pure Pomeron model (without aiming at quantitatively reproducing
the data), and also in a "more realistic" model including Pomeron,
Odderon and secondary Reggeons, fitted to the high energy data for
$pp$ and $\bar pp$ elastic scattering~\cite{dj}. The somewhat
surprising results of this "realistic" approach were:

{\it (i)} a failure to find within the eikonalized model as high
quality a fit as within its Born approximation~\cite{dgj}, even when
readjusting the parameters and even when confining oneself to the ISR
data, limited to low-$|t|$~;

{\it (ii)} a rapid numerical convergence of the rescattering series:
a limited number of rescatterings (4, in addition to the
Born term) is sufficient to obtain a very good approximation at
present energies~;

{\it (iii)} when rescattering corrections were taken into account, a
second break in the slope revealed around $|t|\sim$ 4 GeV$^2$ in the
angular distribution at 300-500 GeV, creating the seed of a
diffraction-type pattern at higher energies~; this break becomes a
shoulder and then a true dip moving down to $|t|\sim$ 3 GeV$^2$ when
$\sqrt{s}$ increases up to 14 TeV. This substructure should be seen
at LHC but might even be detected at RHIC~\cite {exp} if the data are
very precise.

Going one step further, the one-parameter extension (QE) and
much more, the three-parameters generalization (GE) prove very useful
to improve the agreement with the data and, therefore, in removing
the conflict found in {\it (i)} above. In addition, it helps in
understanding the appearance of secondary structures, which stirred
considerable interest and which is intriguing enough that we should
reconsider further both their origin and their model dependence. The
variety of descriptions giving rise to these diffraction-like
multiple structures may suggest them to be essentially model
independent; on the other hand, this is not established in an
unambiguous way and deserves further theoretical analysis\footnote{We
stress once more that several models of $pp$ and $\bar p p$ elastic
scattering (see {\it e.g.}~\cite{dgp,past,dj}) have given hints, in
the past, of the possible appearance of a succession of dips or
shoulders in the angular distributions, at large-$|t|$ values and at
superhigh energies.}. In the light of this, we have undertaken a most
careful analysis of several models both eikonalized and in the Born
approximation, trying to ascertain whether or not the predictions of
secondary structures could be related to some general pattern.
By-products of our investigation turn out to be the verification that
the Odderon intercept $\alpha_O(0)-1\equiv\delta_O$ obtained in the
various fits is invariably non-positive and, empirically, very close to zero and
that the real part of the even
amplitude has the zero predicted by general theorems~\cite{martin} near
$|t|=0$.

In Sec.~2, we report
about several non-eikonalized models with some details on their
specific Pomeron and Odderon components. In Sec.~3, we do the same
about OE, QE and GE. The results are presented in Sec.~4, some
general conclusions are given in Sec.~5.

\section{\bf The input Born }

We focus on the (dimensionless) crossing-even and -odd amplitudes
$a_\pm(s,t)$ of the $pp$ and $\bar pp$ reactions {\footnote{ Here and
in the following, we denote by lower case letters the Born (or input)
amplitudes and by the corresponding capital letters their eikonalized
counterparts.}}
\begin{equation}
a_{pp}^{\bar pp}(s,t)=a_+(s,t)\ \pm a_-(s,t)\ ,
\end{equation}
for which we have data{\footnote{ For all versions, we fitted the
adjustable parameters over a set of $\sim 1000$ $pp$ and $\bar pp$ data of
both forward observables (total cross-sections $\sigma_t$ and
$\rho-$ratios of real to imaginary part of the amplitude) in the
range $4\le\sqrt{s}$ (GeV)$ \le 1800$ and angular distributions
(${d\sigma\over dt}$) in the ranges $23\le \sqrt{s}$ (GeV)$ \le 630$
and $0\le |t|\le 14$ GeV$^2$. The references to the original
literature can be found in~\cite{dgp}.}} on~:

i) total cross-sections
\begin{equation}
\sigma_t = {4\pi\over s}\Im{\rm m} A(s,t=0) \ ,
\end{equation}
ii) differential cross-sections
\begin{equation}
{d\sigma \over dt}={\pi\over s^2}\big|A(s,t)\big|^2 \ ,
\end{equation}
iii) ratio of the real to the imaginary forward amplitudes
\begin{equation}
\rho={\Re{\rm e} A(s,t=0)\over \Im{\rm m} A(s,t=0)} \ .
\end{equation}

The crossing even part in the Born amplitude is a Pomeron (to which a
$f-$Reggeon is added)
while the crossing odd part is an Odderon
(plus an $\omega-$Reggeon)
\begin{equation}
a_+(s,t)=\ a_{P} (s,t)+\ a_f(s,t) \ ,\quad
a_-(s,t)=\ a_{O} (s,t)+\ a_\omega(s,t) \ .
\end{equation}

For simplicity the two Reggeons have been taken in the standard form
\begin{equation}
a_R(s,t)= a_R\tilde s^{\alpha_R(t)}\ e^{b_Rt} ,
\quad \alpha_R(t)=\alpha_R(0) + \alpha'_R t \ ,
\quad (R=f\, {\rm and}\, \omega)  \ ,
\end{equation}
where $a_f$ ($a_\omega$) is real (imaginary). We begin with
trajectories whose parameters are fixed as in previous works (for
example~\cite{dgp}) $\alpha_f(t)=0.69 + 0.84\, t$, and
$\alpha_\omega(t)=0.47 + 0.93\, t$ (with $t$ in GeV$^{2}$), close to
the values obtained in other
recent fits
(\eg ~\cite{c}). As it turns out, however, a best fit requires
some variation of these parameters. Thus, at the price of
economy in the parameters, we end up letting them vary.

We have investigated wide classes of choices where the input
amplitude ("Born term") for the Pomeron $a_{P}(s,t)$ and for the Odderon
$a_{O}(s,t)$
is either a monopole (\ie a simple
pole in the angular momentum $J-$plane) or a "dipole" (\ie a linear
combination of a simple pole with a double pole).

The forms of $a_{P} (s,t)$ in the case of a monopole $(M)$ and of
a dipole ($D$) are
\begin{equation}
a_P^{(M)}(s,t)= a_P \ \tilde s^{\alpha_P(t)} e^{b_Pt}\ ,
\end{equation}
and
\begin{equation}
a_P^{(D)}(s,t)= a_P \ \tilde s^{\alpha_P(t)}
\left[e^{b_P(\alpha_P(t)-1)} (b_P+\ell n{\tilde s})
                                     \ +\ d_P\ell n{\tilde s}\right]\ ,
\end{equation}
where $a_P$ is real.

The difference between a monopole and a dipole results in an
amplitude for the second that grows with an additional power of $\ell
n s$.

 The Odderon may be constructed with the same requirements.
 It is, however, known that the r\^ole of the Odderon at $t=0$ is
negligible but no theoretical prescription is known as how to cut it.
A simple way out is to multiply the monopole or dipole form by a
convenient damping factor. We choose
\begin{equation}
a_O(s,t)=(1-\exp{\gamma t}) a_O^{(M)}(s,t)\ ,
\end{equation}
or
\begin{equation}
a_O(s,t)=(1-\exp{\gamma t}) a_O^{(D)}(s,t)\ .
\end{equation}
In (9) (or (10)), the amplitude on the r.h.s. is constructed along the
same lines as in (7) (or (8)) for $a_P^{(M,D)}(s,t)$. $a_{P}$, however,
is real while
$a_{O}$ is imaginary. As usual,
\begin{equation}
\tilde s\ =\ {s\over s_0} \ e^{-i{\pi\over 2}}\ ,\quad
(s_0=1\ {\rm GeV}^2)\ ,
\end{equation}
enforces $s-u$ crossing and $\alpha_i(t)$ are the trajectories
taken, for simplicity, of the linear form\footnote{Linear
trajectories are an oversimplification that, strictly, violates
analyticity. In addition, at large $|t|$ this may be dangerous in
practice. We ignore this complication.}
\begin{equation}
\alpha_i(t)= 1+\delta_i+\alpha_i't  \  ,(i=P,O)\ .
\end{equation}

It appears impossible to discriminate between
(D) or (M), on general grounds; only the phenomenological
results seem to prefer
(D) over (M). For
the sake of economy we confine our presentation to the dipole case,
which gives somewhat better phenomenological results.

Some authors maintain that a perturbative ({\it a large-$|t|$}) term
behaving like $|t|^{-4}$ (and complying with perturbative QCD
requirements accor\-ding to~\cite{dl}~\footnote{ We should, however,
not forget that at, even at the largest \t values, the ratio $|t|/s$ is
really rather small
so that we are in a domain closer to the usual Regge kinematics than
to that of perturbative QCD. }) is to be added to the Odderon. When
the Born amplitude is eikonalized, however, all rescattering
corrections implied by eikonalization are, in principle, already taken
into
account. Adding another large-$|t|$ term at the Born level would
mimic further rescattering corrections and would lead to double
counting in the eikonalized models. We shall not consider this
option.

We remark that most good fits require $\delta_{P}>0$ implying what is
known as a supercritical Born Pomeron {\it i.e.} a Born amplitude which,
taken
at face value, will eventually exceed the Froissart-Martin~\cite{fm}
unitarity bound even though at extremely high energies (other kinds
of troubles would arise much earlier~\cite{kpp}). This special
violation of unitarity is removed by {\it all} kinds of
eikonalization. Nevertheless, one must verify that the unitarity
constraints
\begin{equation}
\delta_P\ge\delta_O\ , \quad {\rm and}\quad\alpha'_P\ge\alpha'_O\ \ .
\end{equation}
are satisfied (see~\cite{gmp,ffkm}).
The slope parameter for the Pomeron, finally, is expected to be in the
vicinity of its
"world" value $\alpha'_P\simeq 0.25$ GeV$^{-2}$ and this turns out to
be, indeed, the result of the fit (see Section 4.2).

As the last comment, we recall that, in the context of the choice of
the eikonalization procedure, a singular solution is, in principle,
possible, whereby the Odderon dominates over the Pomeron~\cite{gmp}.
For the sake of completeness, we have also tried this option,
unphysical as this appears but, as expected, such possibility is
ruled out by the results of the fits; the fit with an Odderon
dominating over the Pomeron is rather poor and unacceptable.

\section{ Eikonalization procedures}

In eikonal models, the scattering amplitudes are expressed in the
impact parameter ("$b$") representation. First, one defines
the Fourier-Bessel's (F-B) transform
of the Born amplitude
\begin{equation}
h^{\bar pp}_{pp} (s,b)= {1\over 2s}\int_0^\infty
a^{\bar pp}_{pp} (s,-q^2) J_0(bq) q\, dq
\quad {\rm with} \quad            q=\sqrt{-t}\ .
\end{equation}
This is related to the eikonal function ("eikonal" for brevity) by
\begin{equation}
\chi^{\bar pp}_{pp} (s,b)\  =\ 2\ h^{\bar pp}_{pp} (s,b)\ .
\end{equation}
The (complete) analytical forms of the Born amplitudes (both (M) and (D))
in $b$-space are given in Appendix A.

In all eikonalization procedures, one first derives the eikonalized
amplitude $H^{\bar pp}_{pp} (s, b)$ in the $b$-representation;
the inverse F-B transform leads then to the usual eikonalized amplitude in
the $s-t$ space
\begin{equation}\label{16}
A^{\bar pp}_{pp} (s,t)= 2s\ \int_0^\infty
H^{\bar pp}_{pp} (s,b) J_0(b\sqrt{-t}) b\, db\ .
\end{equation}
The main technical problem of eikonalization is the
derivation of $H^{\bar pp}_{pp} (s, b)$ once $h^{\bar pp}_{pp} (s,
b)$ are given. In what follows we make explicit this step in, we
believe, the most general form so far derived.

\subsection{Ordinary and Quasi Eikonalization}

In the ordinary eikonal (OE) formalism, $H^{\bar pp}_{pp} (s,b)$ is
the sum over all rescattering diagrams in the approximation
when there are only two nucleons on the mass shell in any
intermediate state
\begin{equation}\label{17}
H^{\bar pp}_{pp,QE}(s,b)\ =\ {1\over 2i}\left(
\sum\limits_{n=1}^{\infty}
{\big[2ih^{\bar pp}_{pp}(s,b)\big]^n \over n!} \right)\ .
\end{equation}
This limitation neglects the possibility to take
into account multiparticle states.
In the quasi eikonal (QE) procedure~\cite{ter}, the effect of
these multiparticle states
in the
various exchange diagrams
is realized
introducing one additional "weight" parameter $\lambda$ and the
eikonalized amplitude in the $b$-representation (17) is
replaced by
\begin{equation}\label{18}
H^{\bar pp}_{pp,QE}(s,b)\ =\ {1\over 2i}
\sum\limits_{n=1}^{\infty}
\lambda^{n-1} {\big[2ih^{\bar pp}_{pp}(s,b)\big]^n \over n!} \ .
\end{equation}
The above series is meant to
represent the sum of all possible multiple exchanges of Pomerons,
Odderons and secondary Reggeons ($n=1$ corresponds to the Born
approximation, $n=2$ to double exchanges, \etc ). Its explicit
analytical form is
\begin{equation}\label{19}
H^{\bar pp}_{pp,QE}(s,b)\ =\ {1\over 2i\lambda}\left(\exp\big[i\lambda
\chi^{\bar pp}_{pp} (s,b)\ \big]-1\right) \ .
\end{equation}
As it is obvious,
the value $\lambda=1$ corresponds to OE, which appears, therefore, as a
particular case of QE.

\bigskip
However, it is not clear why all intermediate states between
the exchanges of two Pomerons or two Odderons (or between one Pomeron
and one Odderon) could be described by just one and the same parameter
$\lambda$ or, differently stated that all the weights for the various
intermediate internal couplings (two Pomerons, two Odderons or one
Pomeron and one Odderon) should be the same. It would appear more
"natural" that
the various exchanges should
require
different weights.
Differently rephrased, in the QE
procedure, we do not distinguish intermediate states between $P-P,
\,O-O$ and $P-O$ exchanges. Giving up this assumption gives rise to a
new kind of generalized eikonal (GE) procedure where all these
intermediate states may have different weights.

\subsection{Generalized Eikonalization}
\subsubsection{with 3 $\lambda :\ \lambda_\pm,\ \lambda_0.$}

Consider again the separate form of the amplitude (1), and let the
crossing-even and crossing-odd input in the $b$-representation be
\begin{equation}\label{20}
h_\pm \equiv h_\pm(s,b)=\ {1\over 2s}\int_0^\infty \ dq \ q\,
J_0(bq)\, a_\pm (s,-q^2)\ ,\quad (q^2=-t) \ .
\end{equation}
Here, postponing for a moment the consideration of the most general
scheme (5) when secondary Reggeons are included, we temporarily
simplify the notation for the crossing even- and the crossing
odd-part as if they were made by just the Pomeron and the Odderon
respectively (later, we will reinstate the complete contribution)
\begin{equation}
a_+(s,t)=\ a_{P} (s,t) ,\quad
a_-(s,t)=\ a_{O} (s,t) \ .
\end{equation}

{\it A priori}, we have three different configurations of exchanges in
the intermediate states which we show diagrammatically in Fig.~1 and
where the various possibilities, $P-P, \,O-O$ and $P-O$ are
described, phenomenologically, by three constants $\lambda_+,\,
\lambda_-,\ \lambda_0$.

\begin{figure}[ht]
\begin{center}
\includegraphics*[scale=0.55]{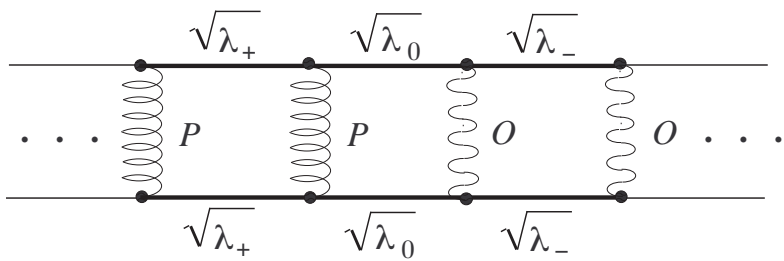}
\caption {An example of a rescattering diagram containing the
possible intermediate states with Pomeron and Odderon exchanges.}
\label{fig.1}
\end{center}
\end{figure}


With this notation, we can deduce
\begin{equation}
 H^{\bar pp}(s,b)\ =\ h_+\ +\ h_- \ +\ H[PP]+H[OO]+\ H[PO]\ +H[OP]\ .
\end{equation}
where (see~\cite{eikge} for the details of the derivation)
\begin{equation}\label{23}
\begin{array}{rll}
2i\lambda_+H[PP]&=
\sum\limits_{n=2}^\infty\sum\limits_{m=1}^{n-1}\sum\limits_{i=1}^{m}
\frac{1}{(m+n)!}{n-1\choose i}{m-1\choose i-1}z^ix^ny^m
\\
\displaystyle &+ \sum\limits_{n=2}^{\infty}\sum\limits_{m=n}^{\infty}
\sum\limits_{i=1}^{n-1} \frac{1}{(m+n)!}{n-1\choose i}{m-1\choose
i-1}z^ix^ny^m + \sum\limits_{n=2}^{\infty}\frac{1}{n!}x^n\\ & = z\
\sum\limits_{n=2}^{\infty}\sum\limits_{m=1}^{\infty} {x^n
y^m\over(n+m)!}\ (n-1) \ _2F_1(1-m,2-n;2;z) \\ &+e^x-x-1\ ,
\end{array}
\end{equation}
\begin{equation}\label{24}
\begin{array}{rll}
 2i\lambda_0H[PO]&=\sum\limits_{n=1}^{\infty}\sum\limits_{m=1}^{n}
\sum\limits_{i=1}^{m} \frac{1}{(m+n)!}{n-1\choose i-1}{m-1\choose
i-1}z^i x^ny^m
\\ &+
\sum\limits_{n=1}^{\infty}\sum\limits_{m=n+1}^{\infty}
\sum\limits_{i=1}^{n} \frac{1}{(m+n)!}{n-1\choose i-1}{m-1\choose
i-1}z^i x^ny^m\\ &=z\
\sum\limits_{n=1}^{\infty}\sum\limits_{m=1}^{\infty} {x^n
y^m\over(n+m)!}\ _2F_1(1-m,1-n;1;z)
\end{array}
\end{equation}
with
$$
x=2i \lambda_+ h_+\ ,\quad y=2i \lambda_- h_-\, \quad
z=\frac{\lambda_0^2}{\lambda_+\lambda_-}.
$$
 $H[OO]$ is obtained from $H[PP]$ with the replacement
$h_+\longleftrightarrow h_-$ and $\lambda_+\longleftrightarrow
\lambda_-$,
and $H[OP]=H[PO]$.

The amplitude $H_{pp}(s,b)$ has the same form as $H^{\bar
p p}(s,b)$ with the replacement $h_-\longleftrightarrow -h_-$.

Unexpectedly,
one can obtain a compact analytical
form from (22). Omitting all details of calculations,
which can be found in~\cite{eikge}, the final expression for
the three-parameters eikonalized amplitudes are
\begin{equation}\label{25}
\begin{array}{rll}
H^{\bar pp}_{pp,GE}(s,b) & =
            \displaystyle {{i\over 2
            (\lambda_0^2-\lambda_+\lambda_-)}}
\bigg\{a + e^{ \displaystyle i(\lambda_+ h_+ \pm \lambda_- h_-)}
\\ & \times
\displaystyle \big[-a \cos{\phi_\pm} + i {c_+h_+\pm
c_-h_-\over\phi_\pm} \sin{\phi_\pm}\big]\bigg\} \ ,
\end{array}
\end{equation}
where we have introduced three constants $a$ and $c_{\pm}$ defined as
\begin{equation}\label{26}
a=2\lambda_0-\lambda_+-\lambda_-\ ,
\end{equation}
\begin{equation}\label{27}
c_\pm=
\lambda_+\lambda_--2\lambda_0^2-\lambda_\pm^2+2\lambda_0\lambda_\pm \
,
\end{equation}
in terms of the parameters of the model and the functions (of $s$ and
$b$)
\begin{equation}\label{28}
\phi_\pm=\sqrt{(\lambda_+h_+ \mp \lambda_-h_-)^2\ \pm 4\lambda_0^2h_+
h_-}\ .
\end{equation}

Considering a general case, when there are no any special relations
between $\lambda_i$, we have found in~\cite{eikge} that the unitarity
inequality
$$
|H^{\bar pp}_{pp,GE}(s,b)|\leq 1
$$
can be satisfied, in general, only if $\delta_{O}\leq 0$ \footnote{The
obvious
inequalities $|h_-|\ll |h_+|$ and $|\Real h_+|\ll |\Imag h_+|,\ \Imag
h_+>0$ , which are valid at high energy, are assumed.}. Two special
cases, namely, $\lambda_0^2=\lambda_-\lambda_+$ (see below) and
$\lambda_+=\lambda_0$ allow $\delta_{O}$ to be positive. However
in all cases unitarity requires the following restrictions
\begin{equation}\label{29}
  \delta_O\leq \delta_P, \qquad \alpha'_{O}(0)\leq \alpha'_{P}(0),
  \qquad \lambda_+ \geq 1/2.
\end{equation}

As anticipated above, it is easy to
prove that these results, obtained in the case of 2 Reggeons (P and
O), hold in the case where 4 Reggeons are grouped 2 by 2 to form a
crossing even ($P+f$) and a crossing odd ($O+\omega$) contribution
with the original definitions (5).

\subsubsection{with 2 $\lambda :\ \lambda_\pm .$ }

A considerable simplification is brought if the factorization
$\lambda_0=\sqrt{\lambda_+\lambda_-}$ is assumed (this is also
treated in great details in~\cite{eikge}). In practice, the main
advantage of this particular case are simplified expressions for
the required amplitudes resulting in a significant gain in computer
time when fitting the data. In this case, the eikonalized amplitude
has the form
\begin{equation}\label{30}
\begin{array}{rll}
H^{\bar pp}_{pp, GE}(s,b)& =\ h_+\pm h_-+\
\left(\frac{h_+\sqrt{\lambda_+}\pm h_-\sqrt{\lambda_-}}
{h_+\lambda_+\pm h_-\lambda_-}\right)^2\\ & \times
\displaystyle\left(\frac{e^{2i(h_+\lambda_+\pm h_-\lambda_-)}-1}{2i}
- (h_+{\lambda_+}\pm h_-{\lambda_-}) \right)\ .
\end{array}
\end{equation}

From unitarity, either
\begin{equation}\label{31}
  \delta_O\leq 0, \qquad \lambda_+\geq 1/2 \qquad \mbox{\rm with} \quad
  \lambda_- \quad \mbox{\rm arbitrary}
\end{equation}
or
\begin{equation}\label{32}
  \lambda_-=\lambda_+\geq 1/2 \qquad \mbox{\rm with} \qquad
0\leq\delta_O\leq \delta_P\ .
\end{equation}
The second case (Eq.(32)) coincides with the previously considered QE
me\-thod.

\subsection{Rescattering series (in $s-t$ space).}
The fact that the eikonalization procedures discussed previously lead
to close analytical forms ((19) or (25)) for the amplitudes $H(s,b)$,
allows us, in principle, to use them in the F-B transform (16) in
order to derive the completely eikonalized physical amplitudes
$A(s,t)$. The compact analytical expressions ((19) or (25)), however,
require a very time-consuming numerical integration. The infinite
expansions ((18) or (23),(24)), on the other hand, can be more
convenient if one has a rapid convergence of the rescattering series.
Fortunately, this condition is fulfilled by both the monopole and the
dipole. These models are, therefore, interesting candidates to
test the number and quality of exchanges necessary to give a final
good accuracy in the calculation of the observables.

To be specific, we rewrite the QE amplitude isolating the Born term
\begin{equation}
{A^{\bar pp}_{pp, QE}}(s,t)\ =\ {a^{\bar pp}_{pp}}(s,t)\ +
\sum_{n=2}^\infty   \lambda^{n-1}\ a^{\bar pp}_{pp; n}(s,t)\ ,
\end{equation}
where from (14) and (16)
\begin{equation}
a^{\bar pp}_{pp; n}(s,t)\ =\ {-i\over n!} s \int_0^\infty\left[2i\
h^{\bar pp}_{pp} (s,b)\right]^n J_0(b\sqrt{-t})b\, db\ .
\end{equation}
Each rescattering term can be calculated analytically only in some
specific cases, for example again in the monopole or dipole models (see
\eg\cite{dj} for the dipole, the monopole calculations are less
involved).
In practice, we find that a
finite number of $\sim 4$ terms is sufficient to insure proper
convergence of the rescattering series ($n\in [2,5]$).

In the GE case, we rewrite the amplitude as
\begin{equation}
{A^{\bar pp}_{pp, GE}}(s,t)\ =\ {a^{\bar pp}_{pp}}(s,t) \ +
\sum_{n_+=0}^\infty\sum_{n_-=0}^\infty a^{\bar pp}_{pp;n_+,n_-}(s,t)\ ,
\end{equation}
where, we have to compare (35) with (23), (24) to obtain the
identification. The analytical expressions for evaluating the double
series are given in Appendix B in the (most involved) case of the dipole
model (Pomeron + Odderon + Reggeons).

In agreement with what we found for QE, the
convergence of the rescattering series for GE is obtained by keeping
only the four first terms ($n_\pm\in [0,1]$).

\bigskip
\section{Results}

As already mentioned, only the results for the dipole model are shown in
what follows.

\subsection{Born input amplitude}
We have satisfied ourselves that the general pattern remains always
the same~\cite{dgp}: a wisely chosen "Born" amplitude can reproduce the
data very well but, depending on this choice, the Pomeron (and the
Odderon) become supercritical and the Froissart-Martin bound is,
in principle, exceeded. At the Born level, secondary structures may or may
not appear; when they do, they are generally due to an additive
contribution to the simple (monopole and dipole) models. For
completeness, given the simplicity of the approach, we give in
Table~1 the parameters of the fit.
Surprisingly, the Odderon intercept equals 1, as recently
claimed~\cite{bartels}. The reader, however, should keep
in mind that this Born approach and its parameters should not be
considered as anything fundamental; they can be used as a shortcut
for giving a reasonable account of the existing data but hardly to
derive general properties.

\medskip
\begin{center}
\begin{tabular}{|l|c|c|c|}
\hline
                &{\bf Pomeron}           &{\bf Odderon }\\
\hline
$\delta_i $       &      0.071         &  0.0\\
$\alpha'_i$ (GeV$^{-2})$       &     0.28             & 0.12\\
$b_i$              &     14.56             &  28.1\\
$a_i$              &     -0.066             & 0.10 \\
$d_i$              &      0.07            & -0.06 \\
$\gamma $(GeV$^{-2})$ & - &  1.56\\
\hline
                & {\bf $f$-Reggeon}      &{\bf $\omega$-Reggeon }\\
\hline $ a_R $ & -14.0 & 9.0 \\
 $ b_R $ (GeV$^{-2})$ &1.64 & 0.38\\
 $ \alpha_R(0) $  & 0.72 & 0.46 \\
 $ \alpha'_R $ (GeV$^{-2})$ & 0.50& 0.50\\ \hline
\end{tabular}
\end{center}

\medskip
Table 1. Parameters of the dipole model fitted at the Born level
(dipole Pomeron $i=P$, dipole Odderon $i=O$ vanishing at $t=0$,
secondary Reggeons $R=f,\, \omega$).

\bigskip

\subsection{Eikonalized models}
A general feature of all eikonalizations models is that, even when
the original Born amplitude exceeds the unitarity limit (remind a
good fit generally requires $\delta_P > 0$), this violation is
removed upon eikonalizing.

We remark that the OE procedure does not change the number of
parameters chosen at the Born level; one parameter ($\lambda$) is
added within the QE procedure and two ($\lambda_\pm$) or three
($\lambda_\pm , \lambda_0$) within the GE procedure. Furthermore, we
 may easily reduce the GE model to the QE model by
setting $\lambda_+=\lambda_-=\lambda_0\equiv \lambda $ and the QE model to
the OE
model by setting $\lambda=1$.

We tested all procedures of eikonalization, either complete or
partial. In the latter case, typically, one may choose not to
eikonalize the Reggeons because they do not induce a unitarity
violation. Whatever the procedure for eikonalizing, we find that the
parameters obtained and the conclusions are qualitatively the same.

From the best fit view point some comments help the reader~:

{\it (i)} the set of experimental data which are very difficult to
reproduce with non vanishing eikonalized dipole Odderon are the
ratios $\rho^{\bar pp}_{pp}(s,t=0)$. This justifies our choice (9-10)
of a Born Odderon input vanishing at $t=0$;

{\it (ii)} leaving the secondary Reggeons parameters free to be
adjusted improves considerably the quality of the fit to the dip
in the ISR energy domain.

\medskip
\subsubsection{Results of the OE and QE fits}
Invariably (and surprisingly), ordinary
eikonalization (OE) leads to a fit which is poorer than in the Born
case but secondary structures emerge.

The QE version of the dipole model improved with respect to OE case
is still poorer than the one obtained at the Born level but one finds
a good reproduction of the data up to and including the dip for $pp$
and the shoulder for $\bar pp$.

In the QE version with fixed trajectories for the secondary Reggeons, we
find a "supercritical" \-Pomeron with $\delta_P\simeq 0.06$ ({\it
i.e.} lower than the value found in~\cite{dl}) and a "critical"
Odderon $\delta_O\simeq -0.03$ as expected. The slope parameter for
the Pomeron $\alpha'_P\simeq 0.25$ GeV$^{-2}$ agrees with the "world"
value, and for the Odderon we find $\alpha'_O\simeq 0.11$ GeV$^{-2}$.
The single parameter characterizing the method of quasi
eikonalization with respect to the ordinary one is found closed to
its lower unitarity limit $\lambda\sim 0.5$. This, in practice,
tends to reduce
the effect of high multiple exchanges.

Concerning the shape of the diffraction like-structures, we find
significant modifications due to QE with respect to previous
work~\cite {dj} in which the OE method had been used. Specifically,
the dip-bump secondary structure shifts towards
somewhat
lower-$|t|$ and
delays its appearance till higher energies are reached. More
precisely, in the QE ({\it i}) the first dip moves down from
$|t|\sim$ 1.2 GeV$^2$ to 0.5 GeV$^2$ when $\sqrt s$ goes up from 60
GeV to 14 TeV; ({\it ii}) a break in the slope appears around
$|t|\sim $ 4.0 GeV$^2$ when the energy is around 500 GeV, becoming a
shoulder and then a true dip which recedes to $|t|\sim$ 1.5 GeV$^2$
when $\sqrt s$ increases to 14 TeV.

It is very instructive to compare the relative virtues of OE and
QE. Generally speaking,
as repeatedly stated,
both eliminate
conflicts with
the unitarity limit
and the
convergence of
the rescattering
series is comparable
(see above), but the QE method appears to cure some undesirable
features of the OE, regarding the quality of the fit.

\smallskip
\subsubsection{Results of the GE fits}
Following the same motivations as above for comparing the QE /OE
versions, we discuss now the implications of generalizing the
eikonalization with the two or three parameters $\lambda_\pm
,\lambda_0$ (instead of a unique parameter $\lambda$ for the QE
case), using the same Born amplitude; again, we report only the
dipole results.

 The GE version with two $\lambda$-parameters (and with fixed
Reggeon trajectories) leads to a good reproduction of the data with
well structured secondary dips. The various values of the parameters
are slightly different from the version with one $\lambda$, in
particular $\lambda_+\simeq 0.5$ and $\lambda_-\simeq 0.44$.

\begin{center}
\begin{tabular}{|l|c|c|c|}
\hline
$\lambda_+$       &     0.5                &                   \\
$\lambda_-$       &     0.55                 &                \\
$\lambda_0$       &     1.24                 &                  \\
\hline
                &{\bf Pomeron}           &{\bf Odderon }\\
\hline
$\delta_i $       &      0.073            & -.0.005 \\
$\alpha'_i$ (GeV$^{-2})$  0.27     & 0.054                 & \\
$b_i$              &         9.0         &  26.6\\
$a_i$              &        -0.114          & -0.019 \\
$d_i$              &          0.165        &-0.09  \\
$\gamma $(GeV$^{-2})$ & $-$ & 1.37 \\
\hline
                & {\bf $f$-Reggeon}      &{\bf $\omega$-Reggeon} \\
\hline
$ a_R $    -12.95     & 16.44                 &                 \\
$ b_R $  (GeV$^{-2})$        & 1.24                 & 3.50                 \\
$\alpha_R(0)$         &     0.81             &  0.47               \\
$\alpha'_R$  (GeV$^{-2})$        &          1.07        &  0.57          \\
\hline
\end{tabular}
\end{center}

\medskip
Table 2. Parameters of the dipole model fitted with the more
sophisticated GE procedure (see also Table 1).

\bigskip

The version with three $\lambda$-parameters and fixed trajectories
for the secondary  Reggeons gives also a good reproduction of the data. The
situation about the diffractive structures is partially different, now the break
around $|t|\simeq 4$ GeV$^2$ becomes a dip which moves to $|t|\simeq
3$ GeV$^2$ at TeV energies. The various parameters are close to those
of the previous case (for two $\lambda$-parameters);
the value of $\delta_O$ remains negative and
moves closer to zero.
For the $\lambda$-s we find $\lambda_+\simeq 0.5,\, \lambda_-\simeq
0.1,\, \lambda_0\simeq 0.86$.

The best fit is obtained if we allow some variation for the
intercepts and slopes of the Reggeon trajectories. The values of free
parameters are given in Table 2.

Three points are worth emphasizing:

{\it i)} the Odderon intercept
$\alpha_O(0)-1\equiv \delta_O$ consistently turns out to be negative
in agreement with general arguments~\cite{wos}; however in practice
such a small value is obtained that we do not contradict the most recent QCD
value~\cite{bartels}

{\it ii)} the real part
of the even amplitude automatically exhibits a zero at small \t
values (typically, $|t| \simeq .30$ GeV$^2$ at $\sqrt{s}=546$ GeV)
and this value recedes towards zero as $\sqrt{s}$ increases
(typically $|t|\simeq .27$ GeV$^2$ at $\sqrt{s}=1800$ GeV) and we
predict it at $|t|\simeq .23 $
at $\sqrt{s}=14$ TeV. This result is
in agreement with a general theorem by A.~Martin~\cite{martin}. Also, a
 second zero appears for $|t|\sim 1.5$ GeV$^2$ at $\sqrt{s}=546$ GeV
which moves to $|t|\sim 1.25$ GeV$^2$ at $\sqrt{s}=1.8$ TeV (see Table 3);

{\it iii)} the eikonalized Odderon contributes to reproduce perfectly
the large-$|t|$ region.

\medskip

\begin{center}
\begin{tabular}{|l|c|c|c|}
\hline
{\bf energy}             &{\bf 1$^{st}$ zero}           &{\bf 2$^d$ zero}\\
                          & (GeV$^2$)  & (GeV$^2$)\\
\hline
  546 GeV    &      0.30     &  1.5 \\
1800 GeV     &      0.27         & 1.25  \\
14 TeV    &         0.23       &  0.95\\
40 TeV    &      0.17    & 0.85\\
\hline
\end{tabular}
\end{center}

\medskip
Table 3. Positions ($|t|$ values) of the first two zeros of the real part
of the even eikonalized amplitude.
\bigskip

We observe the same evolution of the structures as in the previous
case.

Concerning the values of the GE parameters, collected in Table 2, we
find a "supercritical" Born \-Pomeron with $\delta_P\simeq 0.073$ ({\it
i.e.} greater
than the QE value), while the slope parameter for the Pomeron is
$\alpha'_P\simeq 0.27$ GeV$^{-2}$ and for the Odderon
$\alpha'_O\simeq 0.05$ GeV$^{-2}$. The tree parameters characterizing
the effect of the generalized method of eikonalization are
$\lambda_+\sim 0.5$,  $\lambda_-=0.55$ and $\lambda_0=1.24$.

Thus, the generalized eikonalization procedure gives better
results than the quasi eikonalization and {\it a fortiori} than the
ordinary eikonalization.

That the $\chi^2/d.o.f.\ (\simeq 7.0)$ remains pretty large is the
consequence of not having made any "wise selection" of the data. The
resulting curves, however, are quite satisfactory as it is shown in
Figs.~2 - 5 (the results are given for the complete set of parameters
in Table~2).

The extrapolations of the total cross section and of the $\rho$-ratio are shown
in Fig.~6. The angular distributions for the energies to be reached in the near
future~\cite{exp} exhibit the secondary structure especially at LHC in Fig.~7.


\vskip 0.4cm
\begin{center}
\includegraphics*[scale=0.44]{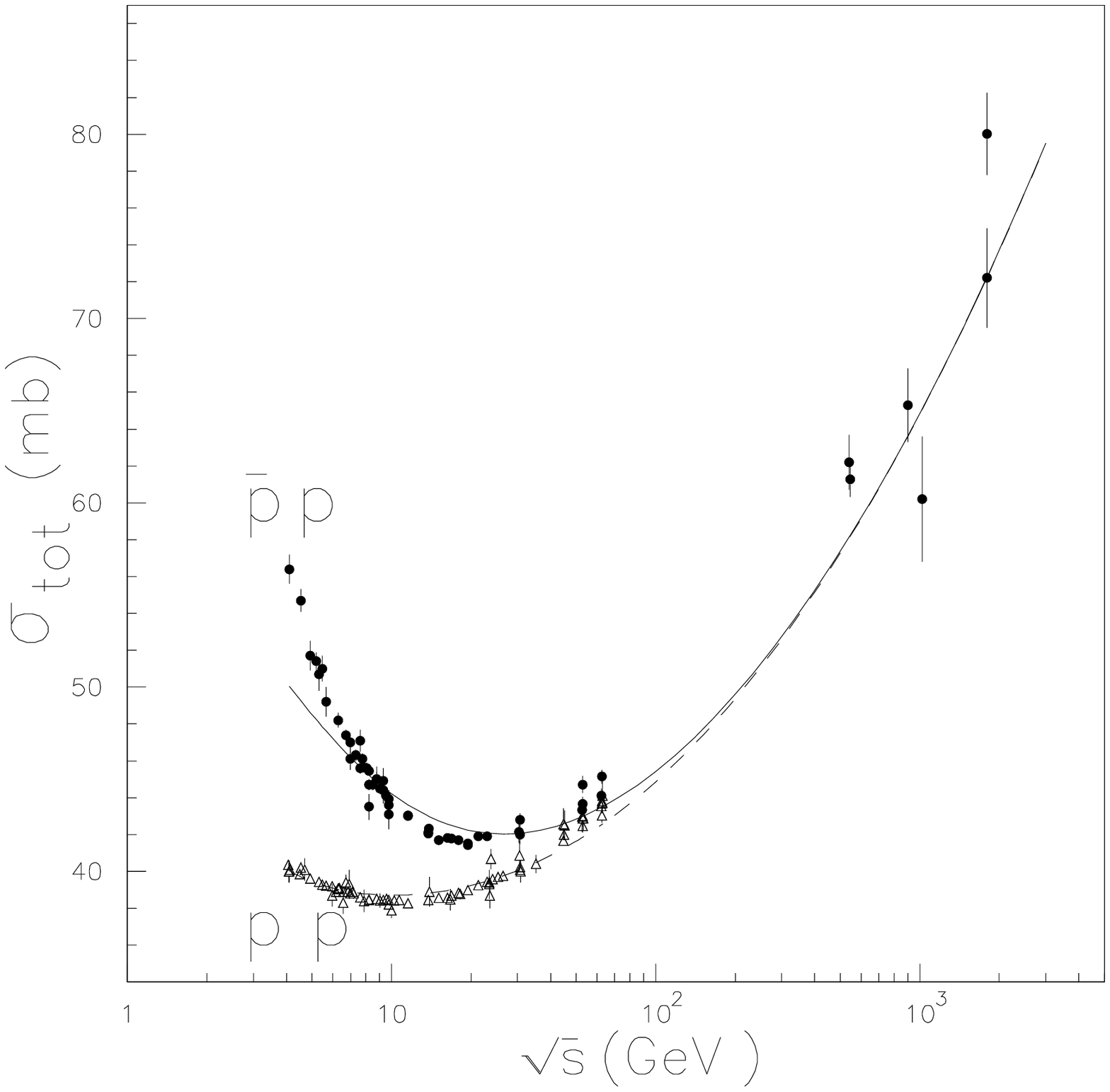}
\end{center}
{\bf Figure 2.} Comparison with the data of the fit to total
cross-sections for $\bar pp$ (full dots) and $pp$ (hollow triangles)
processesfor the most sophisticated generalized eikonalization (GE) procedure.
\vskip 0.4cm
\begin{center}
\includegraphics*[scale=0.44]{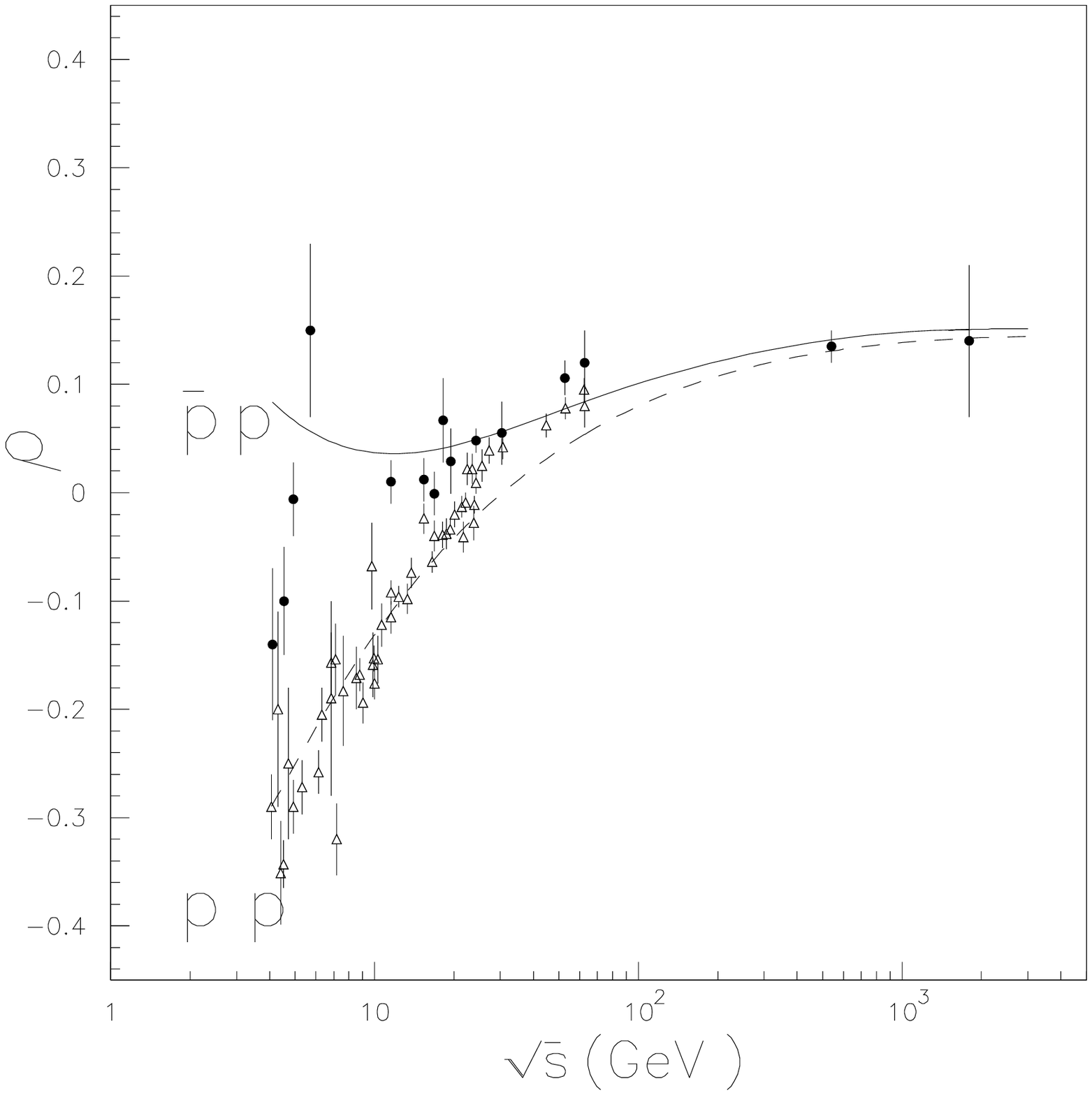}
\end{center}
{\bf Figure 3.} Same as Fig.2 for $\rho$-ratios.


\vskip 0.4cm
\begin{center}
\includegraphics*[scale=0.44]{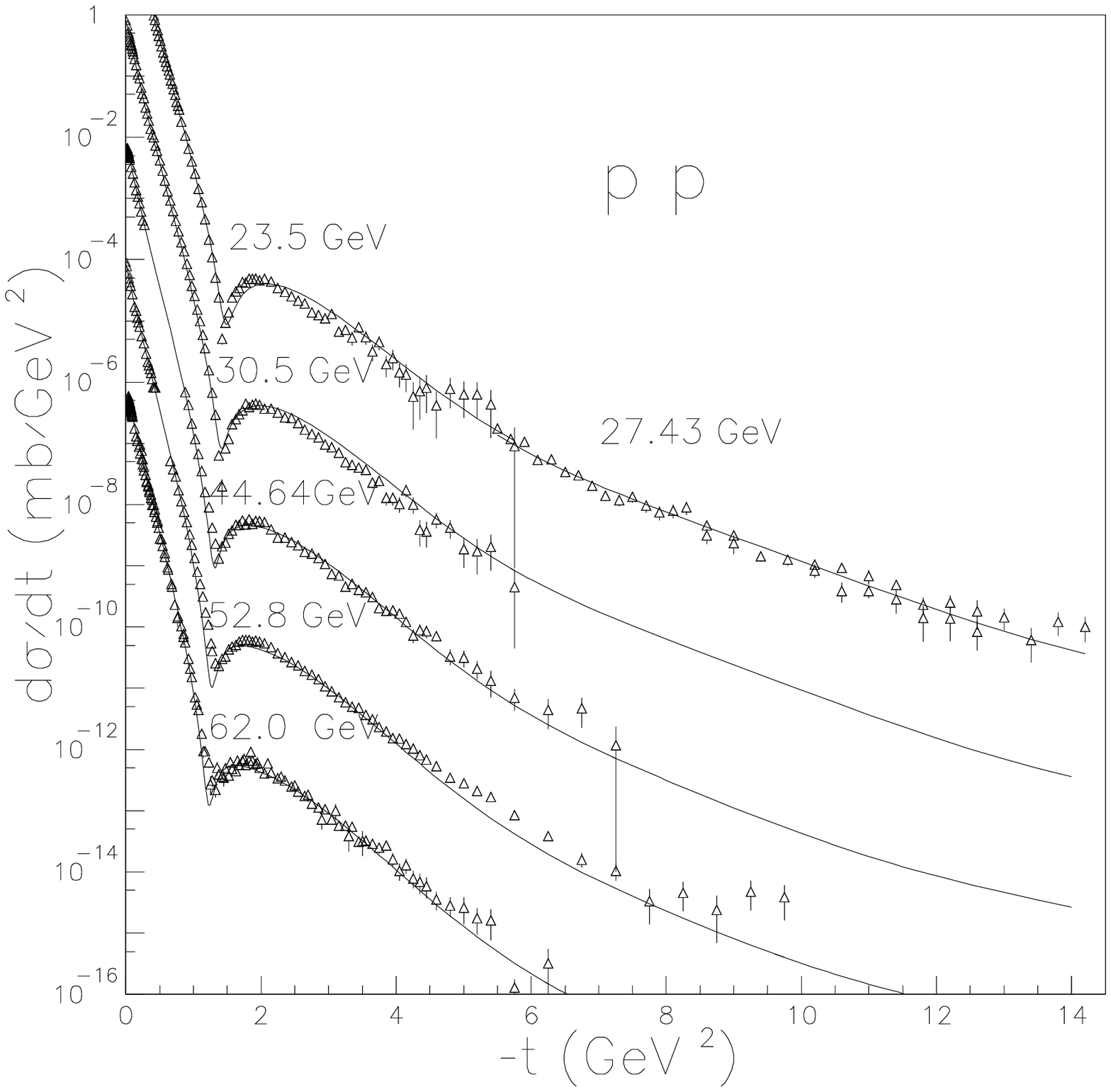}
\end{center}
{\bf Figure 4.} Comparison with the data of the fit to differential
cross-sections for $pp$ process
for the most sophisticated generalized eikonalization (GE) procedure. A
$10^{-2}$ factor between each successive curve is omitted.
\vskip 0.4cm
\begin{center}
\includegraphics*[scale=0.44]{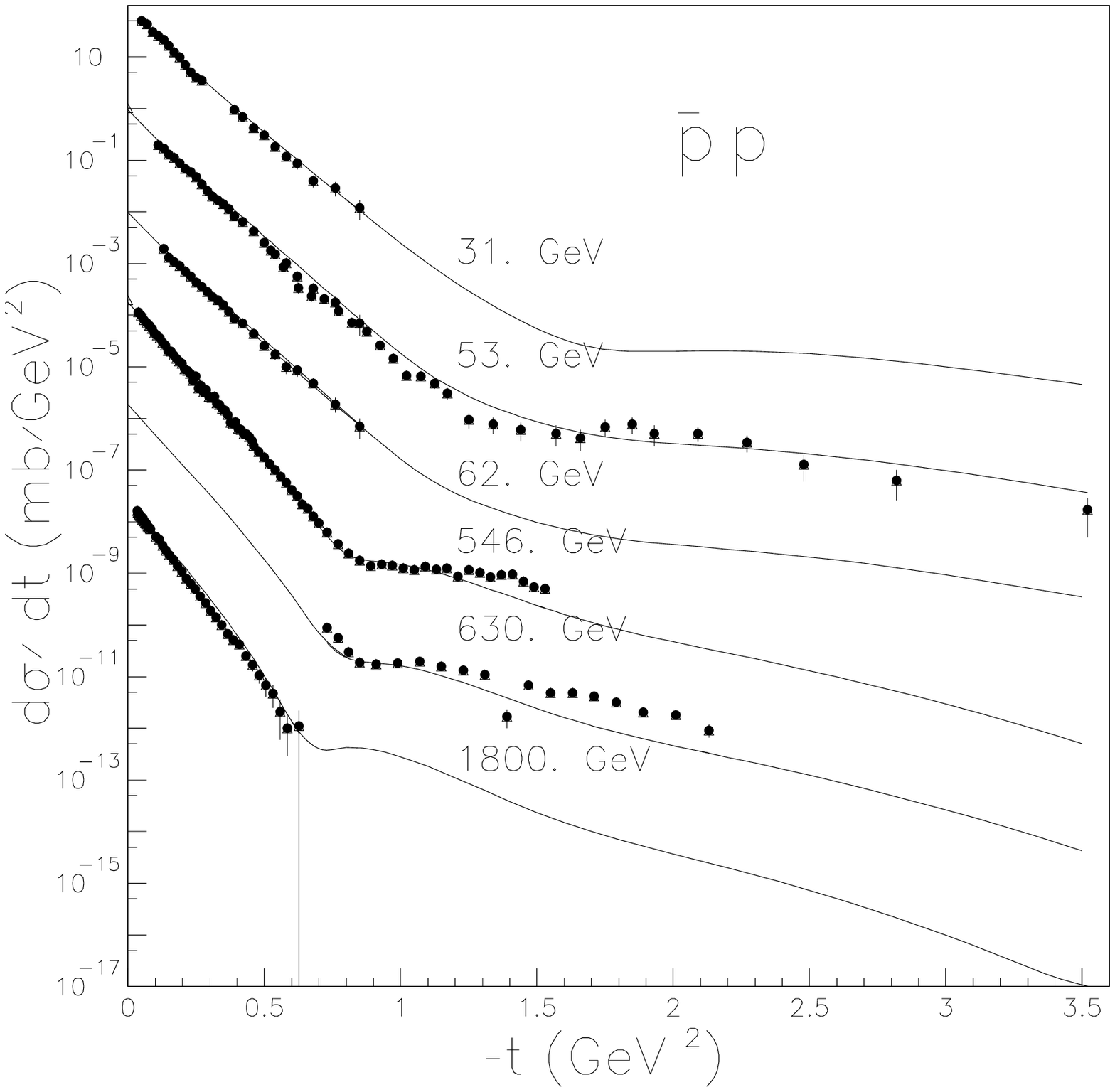}
\end{center}
{\bf Figure 5.} Same as Fig.4 for $\bar pp$ process. The Tevatron data are not
fitted.


\vskip 0.4cm
\begin{center}
\includegraphics*[scale=0.65]{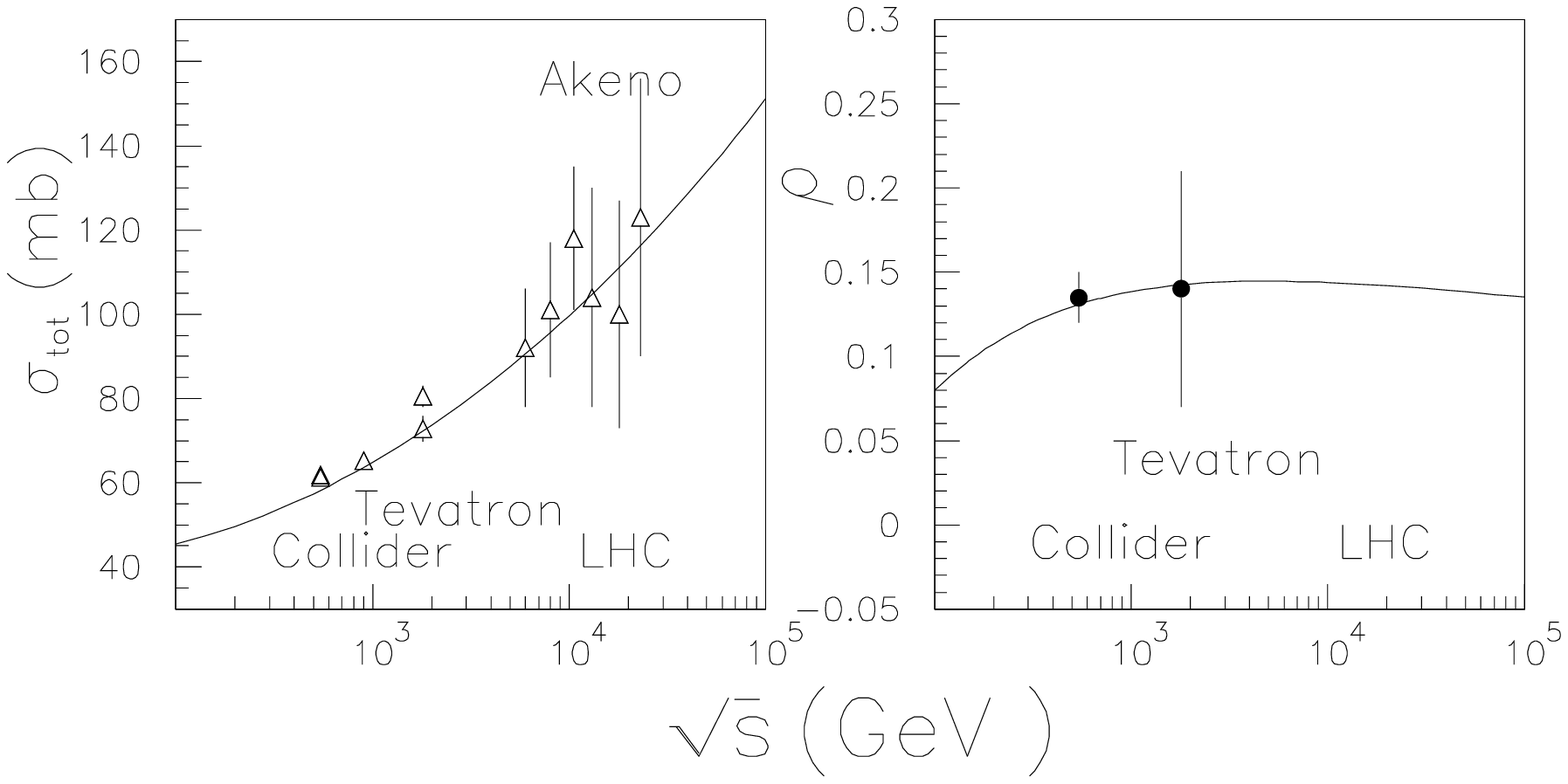}
\end{center}
{\bf Figure 6.} Calculated observables within the GE dipole model,
versus the energy and compared to the data (cf~\cite{dgp})~:
total cross-section $\sigma_{tot}$ (the cosmic ray data are not
fitted) and $\rho$-ratio.
\vskip 0.4cm
\begin{center}
\includegraphics*[scale=0.5]{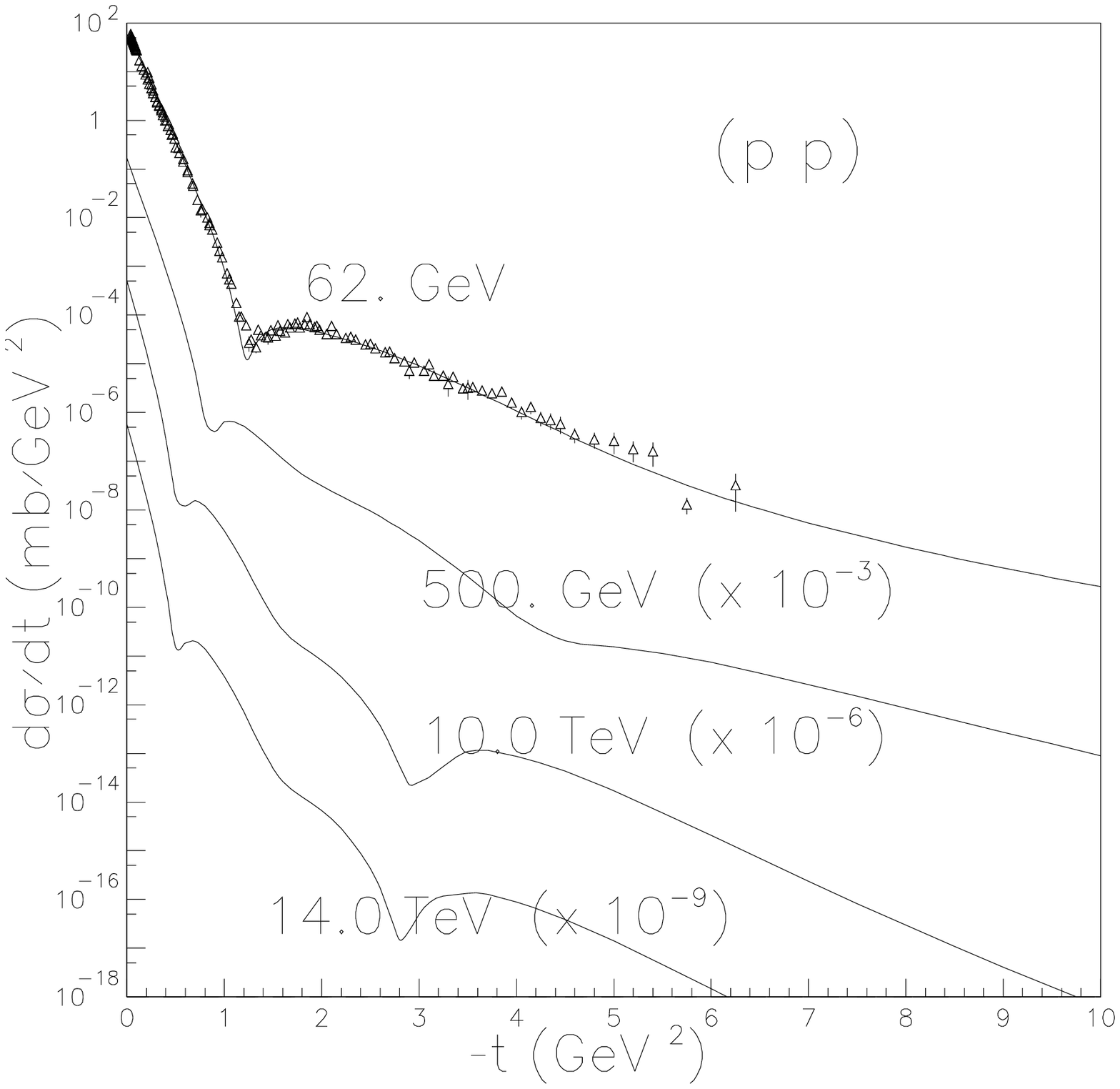}
\vskip 0.3cm
\end{center}
{\bf Figure 7.} Extrapolations to RHIC and LHC energies of the
calculated $pp$ differential cross-sections.

\section{Concluding remarks}
Let us try to answer some of the questions raised in the
Introduction. Of course, we do not have {\it the final prescription}
for the Pomeron. Many of the forms discussed above give a good
reproduction of the data; several of them (and many others in the
literature) seem to work well both at the Born and at the eikonalized
level (in particular, the Dipole Pomeron).

Often, the Born Pomeron is found to be supercritical ($\delta_P > 0$)
which implies an intrinsic problem with unitarity; this is removed by
(all kinds of) eikonalization. Thus, the r\^ole of eikonalization is
very important for the asymptotic behavior of
all physical quantities. In all cases the eikonalization restores the correct
high energy behavior of the supercritical Pomeron. While the data
for total cross-sections
do not contradict the $\ell n^2s $ behavior resulting from the
eikonalization of a supercritical Born Pomeron, they are not
incompatible with a $\ell n s$ form.
The inclusion in the fit of the data at $t\neq 0$
is absolutely necessary to get
an unambiguous conclusion on the behavior of
all physical quantities.

The presence of the Odderon contribution, as repeatedly emphasized,
is necessary to reproduce well the angular distributions data in the
dip-region and for large-$|t|$ values
but its contribution is required by the fit to be
negligibl in the forward domain.

The problem of the Odderon intercept remains very complicated but the
general agreement, in LLA, is now~\cite{gln,wos} that the
Odderon intercept is closed to 1 with $\delta_O < 0$ or
$\delta_O = 0$~\cite{bartels}.
This agrees with our findings (see also~\cite{eikge}).

A burning question concerns whether or not it is possible to get a
definite prediction about the existing of secondary structures. At
the Born level, the presence or absence of secondary structures rests
on the specific properties of the Born amplitude (like an oscillatory
component in the Pomeron amplitude). In this case, therefore, the
prediction of secondary structures appear quite model dependent.
The r\^ole of eikonalization is very important in this context. In
the dipole case, structures appear in the angular distribution as
soon as a double Pomeron exchange is taken into account; the trend
consolidates when the number of rescattering corrections $n$
increases and takes a definite form when
several
exchanges are included. This appears to be the case in
all eikonalization procedures. We conclude that secondary structures
are unambiguously predicted by any eikonalization process. This
reinforces previous conclusions by other authors~\cite{past}. In fact, as
emphasized by Horn and Zachariasen~\cite{hz}, oscillations in
$t$ should be expected from the properties of Bessel
functions in the F-B transforms unless some special feature of the
eikonal destroys them.

Of all eikonalization procedures discussed, GE with 3 parameters
leads to the best account of the data.

Finally, we emphasize that the real part of the even amplitude at high
energy has a zero in the small-$t$ region,
as anticipated by a general theorem~\cite{martin}.

In conclusion, while we believe that LHC will definitely prove (or
disprove) the validity of our predictions of secondary structures and
about the zero of the real part of the even amplitude, we insist on
how valuable it would be to have both $pp$ and $p\bar p$ options
available, at the same machine and at the highest energies in order
to check not only our predictions but a whole host of theoretical
high energy theorems.
\bigskip

{\large{Acknowledgements.}} We would like to thank G. Lamot for
his help with the fortran code in particular
for the hypergeometric functions. Two of us (EM
and EP) would like to thank the Institut de Physique Nucl\'eaire de
Lyon for the hospitality and two of us (MG and EM) would like to
thank the Theory Physics Department of the University of Torino for
the hospitality. Financial support by the INFN and the MURST of Italy
and from th IN2P3 of France is gratefully acknowledged.

\bigskip
\centerline {\bf APPENDIX A}
\bigskip

\centerline{\large {Analytical Born amplitude in the $b$-space}}
\medskip

We have
now to
define
the analytical expressions
of the Born amplitudes in b-space
$$h^{\bar pp}_{pp} (s,b)\ =
\ h_f(s,b)+h_P(s,b)\ \pm\
\left[h_O(s,b)+h_\omega(s,b)\right]\ \equiv h_+\pm h_-\ \eqno(A1)$$
from which
we will derive
the eikonalized amplitude.
With our choices of Born (s,t) amplitudes, all
the analytical F-B's transforms are readily obtained{\footnote{
Recall that the "couplings" $a_f,a_P$ are real and $a_\omega ,a_O$
are imaginary}}; for the secondary Reggeons
$$
h_{R}(s,b)={1\over2}a_R{\tilde s^{\alpha_R(0)}\over s}\
{\exp({-{b^2}\over 4B_R})\over 2B_R}; \quad  B_R=
\alpha'_R \ell n\tilde s+b_R \ , \quad R=(f,\omega ) \ ,\eqno(A2)$$
where we have defined $B_R$ in terms of the slopes $b_R$ introduced
earlier in (6)). The Pomeron part depends on our choice (7 or 8): for
the monopole we would have
$$
h_{P}^{(M)}(s,b)={1\over 2} a_P {\tilde s^{\alpha_P(0)}\over s}\
{\exp({-{b^2}\over 4B_P})\over 2B_P}; \quad B_P= \alpha_B' \ell
n\tilde s+b_P \ ,\eqno(A3)
$$
while for the dipole
$$
h_{P}^{(D)}(s,b)={-i\ a_P\over 4 \alpha'_P s_0}\ \big(
e^{r_{1,P}\delta_P -{{b^2}\over 4B_{1,P}}}\ +\ d_P\ e^{r_{2,P}
\delta_P -{{b^2}\over 4B_{2,P}}} \big)\ .\eqno(A4)
$$
For our Odderon monopole (9) we have
$$
h_{O}^{(M)}(s,b)={1\over 2} a_O {\tilde s^{\alpha_O(0)}\over s}\
\left [{\exp({-{b^2}\over 4B_O})\over 2B_O}- {\exp({-{b^2}\over
4\widetilde{B}_O})\over 2\widetilde{B}_O}\right ],\eqno(A5)
$$
where $B_O= \alpha_O' \ell n\tilde s+b_O$ and $\widetilde{B}_O=
\alpha_O' \ell n\tilde s+b_O+\gamma$. Finally, for
our Odderon dipole (10)
$$
\begin{array}{lll}
h_{O}^{(D)}(s,b)&=&\displaystyle {-i\ a_O\over 4 s_0}\ \Bigg (
e^{r_{1,O}\delta_O -{{b^2}\over 4D_{1,O}}}{r_{1,O}\over D_{1,O}}\ -
e^{r_{1,O}\delta_O -{{b^2}\over 4\widetilde{D}_{1,O}}}{r_{1,O}\over
\widetilde{D}_{1,O}}\\ &+&\displaystyle d_O\ e^{r_{2,O} \delta_O
-{{b^2}\over 4D_{2,O}}}{r_{2,O}\over D_{2,O}}- d_O\ e^{r_{2,O}
\delta_O -{{b^2}\over 4\widetilde{D}_{2,O}}}{r_{2,O}\over
\widetilde{D}_{2,O}} \Bigg )\ .
\end{array}
\eqno(A6)
$$
We have defined
$$
r_{1,i}=\ell n\tilde s+b_i\ , r_{2,i}=\ell n\tilde s\ , (i=P,O)\ ,
\eqno(A7)
$$
and
$$ B_{i,P}= \alpha'_P r_{i,P} \ ,\quad D_{i,O}= \alpha'_O r_{i,O} \ ,
\quad \widetilde{D}_{i,O}= \alpha'_O r_{i,O} +\gamma \ , (i=1,2)\ .
\eqno(A8)$$

\bigskip
\centerline {\bf APPENDIX B}
\bigskip

\centerline{\large {GE Dipole Model and Rescattering series in $s-t$
space}}
\medskip

As mentioned in the text, the monopole and the dipole model are
useful
to study
various properties, such as convergence of the
rescattering series expansion, together with the effect of
generalizing the eikonalization since each rescattering term is
tractable analytically. Here, we
consider
only the dipole case as an example.

The OE dipole model has been investigated in~\cite{dj}.
The extension to the QE case is straightforward. We rewrite the GE
amplitude as
$$
A^{\bar pp}_{pp} (s,t)= 2s\ \int_0^\infty H^{\bar pp}_{pp} (s,b)
J_0(b\sqrt{-t}) b\, db\ , \eqno(B1)
$$
with
$$
H^{\bar pp}_{pp} (s,b) =h_+\pm h_- +H[PP]+H[OO]+2H[PO]\ ; \eqno(B2)
$$
the rescattering contributions $H[PP,OO,PO]$ are given in (23),(24).
We split the Born contribution and the rescattering series of the GE
dipole model (with 3 $\lambda$'s) which runs over the two indexes
$n_\pm$ from 0 to infinity
$$
{A^{\bar pp}_{pp, GE}}(s,t)\ =\ {a^{\bar pp}_{pp}}(s,t) \ +
\sum_{n_+=0}^\infty\sum_{n_-=0}^\infty a^{\bar pp}_{pp;n_+,n_-}(s,t)\
.\eqno(B3)
$$
Introducing the 4 partial contributions of the eikonal function $\chi
(s,b)$
$$
h_+ = \ {1\over 2}\ \left(\chi_P(s,b)+\chi_f(s,b)\right) \quad
  h_- = \ {1\over 2}\ \left(\chi_O(s,b)+\chi_\omega(s,b)\right)
  \ ,\eqno(B4)
$$
known analytically from
Appendix A
and separating the three contributions, we obtain in the GE dipole case
$$
a^{\bar pp}_{pp;n_+,n_-}(s,t)=
\ i\ s\ {(\pm i)^{n_++n_-}(\lambda_+)^{n_+}\ (\lambda_-)^{n_-}
\over (n_++n_-+2)! }
$$
$$
\times \ \left(F_{n_+,n_-}. \displaystyle{I}+F_{n_-,n_+}.
\displaystyle{II}+ G_{n_+,n_-}. \displaystyle{III}\right) \ ,
\eqno(B5)
$$
where we have introduced the hypergeometric functions $_2 F_1$ (with the real
argument $z={\lambda_0^2\over \lambda_+ \lambda_-}$)
$$
F_{n_\pm,n_\mp} = z (n_\pm+1). _2 F_1(1-n_\mp,-n_\pm;2;z).
(1-\delta_{n_\mp,0}) +
\delta_{n_\mp,0} \ ,
$$
$$
G_{n_+,n_-} =_2 F_1 (-n_-,-n_+;1;z)\ .
$$
In (B5) we have also defined the inverse F-B's transforms
$$
 I  =\ \lambda_+ \sum _{\ell =0}^{n_++2} \sum_{m=0}^{n_-}
\pmatrix{n_++2\cr \ell\cr}\pmatrix{n_-\cr m\cr} {\rm
Int}_{n_++2-\ell, n_--m,\ell ,m} (s,t)\ , \eqno(B6)
$$
$$
 II  =\ \lambda_- \sum _{\ell =0}^{n_+}\sum_{m=0}^{n_-+2}
\pmatrix{n_+\cr \ell\cr}\pmatrix{n_-+2\cr m\cr} {\rm Int}_{n_+-\ell,
n_-+2-m,\ell ,m} (s,t)\ , \eqno(B7)
$$
$$
 III  =\ \pm 2{\lambda_+\lambda_-\over\lambda_0}
\sum _{\ell =0}^{n_++1}\sum_{m=0}^{n_-+1} \pmatrix{n_++1\cr
\ell\cr}\pmatrix{n_-+1\cr m\cr} {\rm Int}_{n_++1-\ell, n_-+1-m,\ell
,m} (s,t)\ . \eqno(B8)
$$
Once again, in these expressions $+ (-)$ corresponds to $\bar pp$
($pp$); $\pmatrix{n\cr k\cr}$
is the binomial c\oe fficient and Int $(s,t)$ is the following integral
over the 4 components of the eikonal function
$$ {\rm Int}_{\lambda,\mu,l,m}\ (s,t)=\int_0^\infty
\chi_P^\lambda(s,b)\,\chi_O^\mu(s,b)\,\chi_f^l(s,b)\,
\chi_{\omega}^m(s,b)\;
J_0(b\sqrt{-t})b\, db. \eqno(B9)
$$
An analytic expression for this
integral has been written~\cite{dj} in the case when the Odderon does
not contain a killing factor at $t=0$. It is
a straightforward
exercise to derive the complete analytical
form from (B9).


\end{document}